# The organising vision of integrated health information systems

Gunnar Ellingsen

University of Tromsø (UiTø)

gunnar.ellingsen@unn.no

and

Eric Monteiro

Norwegian University of Science and Technology (NTNU)

eric.monteiro@idi.ntnu.no

**Abstract**

The notion of 'integration' in the context of health information systems is ill-defined yet in wide-spread use. We identify a variety of meanings spanning from purely technical integration of information systems to integration of services. This ambiguity (or interpretative flexibility), we argue, is inherent rather accidental: it a necessary prerequisite for mobilising political and ideological support among stakeholders for integrated health information systems. Building on this, our aim is to trace out the career dynamics of the vision of 'integration/ integrated'. The career dynamics is the transformation of both the imagery and material (technological) realisations of the unfolding implementation of the vision of integrated care. Empirically we draw on a large, ongoing project at the University hospital of North Norway (UNN) to establish an integrated health information system.

## 1. Introduction

The health care sectors in all Western countries are profoundly fragmented. They are fragmented across geographical, professional, organisational and technical boundaries. This creates a fragmented health care *service* for the patients (SHD, 2007; Boochever, 2004), which undermine efforts of transforming organizations towards more collaborative, process-oriented modes of working (Broadbent et al., 1999; Scott-Morton, 1991). Ultimately, this threatens ongoing efforts to increase the efficiency and effectiveness of health care delivery.

This generates the very understandable motivation for notions like 'continuity of care', 'shared care' or '*integrated* care'. 'Integration' is a potential remedy for problems fragmentation produce, i.e. duplication, non-consistency and redundancy. To appreciate the importance of integrated information systems, Davenport (1998:127) explains, we "need to understand the problem [they] are designed to solve: the fragmentation of information in large …organizations". Visions of 'integrated care' are attractive exactly because they promise a solution to these concerns including efficiency gains and quality improvements.

Despite its common use and perceived attractiveness, however, the qualifier 'integration/ integrated' in the context of information systems remains notoriously slippery as "individuals often have a different understanding of the meaning of the word" (Gulledge, 2006:5). It marks the political and ideological commitment towards integrated care, i.e. service integration as experienced by individual patients. On the other hand, it is simultaneously used in the much narrower sense of the technical integration of relevant clinical information systems.

We describe the *distinct meanings* of the qualifier 'integrated/ integration' in relation to health information systems and analyse its *ambiguity*. We argue that the implementation process of integrated systems in health care rely on the ambiguous meaning of 'integration' i.e. necessary support for the project is due to, not despite, the ambiguous meaning of 'integration'. Identifying ambiguity – the interpretative flexibility whereby different actors or stakeholders ascribe distinct meanings to an artefact or concept – should hardly surprise anyone. Our ambition is to go beyond the mere recording of this ambiguity. The principal aim of this paper is to analyse the *career* or sequence of transformations the vision of integrated health information systems undergo as it interacts with the intended as well as unintended consequences of its implementation. In other words, we analyse the reciprocal transformations over time of the vision of integrated care and its material (attempted) realisations among involved stakeholders.

Empirically we draw on a large, ongoing project at the University hospital of North Norway (UNN) to establish an integrated health information system. Born out of increasing impatience and frustration with the exiting fragmented portfolio, consisting of an electronic patient record system (EPR) (which we dub RecSys), radiology-system (RIS), patient administrative system (PAS) and a fair share of the laboratory-systems all provided by different vendors, the intention was to replace this, non-integrated portfolio with a tightly integrated package from one vendor which we refer to as HealthSys.

## 2. Conceptualising integration in health care: visions and technologies

Policy documents on Norwegian, not unlike other European, health sector reforms, rely heavily on ICT to meet the serious challenges for increased efficiency, effectivness and improved quality of the whole value chain of health care services. Integration, a key policy document argues, implies that all relevant information would be always available:

> "The health care worker becomes more available because all information is accessible regardless of where the health care worker and patient are" (SHD, 2001)

It goes on to illustrate how integration facilitates the creation of a health service virtual market where:

> "In the future there will probably be an increase of services in 'networks' where bidders can present services [for example on the internet] and those who want, for instance referral physicians, can search for services, check capacity and, if possible, make bookings. [Accordingly] the network will be a trading place" SHD (2001).

In the most recent policy document from the Norwegian directorate for Social and Health affairs, the crucial role of integrated information systems to achieve integrated (or continuity of) care is emphasised:

> "The aim is that patients and clients shall experience continuity of care when they use the [health] services. Electronic interaction is decisive in order to ensure the free flow of information" (SHD, 2007:5)

Even this small sample of the meanings of the vision of 'integration' in the health care sector demonstrates its varied meanings, thus ambiguity: flexible access of information, creation of virtual markets of services with their consumption or continuity of care.

Rather than being surprised or confused by this, we need to recognise that the ambiguity over the exact meaning of 'integrated care' is typical for all visions. A characteristic, not to say *constitutive*, feature of visions are their ambiguity. A highly relevant analysis of this aspect of visions is Swanson and Ramiller (1997) who introduce the notion of 'organizing vision'. Swanson and Ramiller (ibid.) explain the productive roles the ambiguity of effective visions play. A key part of a vision is to reduce uncertainty during planning and decision making. They underscore the vision's ability to provide multiple interpretation and "institutional coherence to initiatives" (ibid: 460). More interestingly, they move beyond simply recognising the ambiguity of visions. They point out the necessity of visions to evolve over time in such a way as to ensure *legitimacy* and:

> "An organizing vision often grounds the legitimation of the [vision] in broader business concerns, linking the innovation to aspects of business functioning that are currently of prominent interest." (ibid: 461).

Finally, the organising vision serves to *mobilise* the necessary resources to support the realisation of the vision as it serves to "activate, motivate, and structure…the innovation (ibid:461).

It is vital to note that the vision do transform; it is not just a question of forging an alliance (legitimation, mobilisation) of different meanings or interpretations. This aspect of a vision is highlighted in Knorr Cetina's (2001: 182) notion of 'epistemic object' and their "changing, unfolding character – or [their] lack of 'object-ivity' and completeness of being". Epistemic objects or visions are thus not objects of which we have complete, potentially diverging, knowledge; they are essentially incomplete and discovered over time in interaction with practice.

So much for visions. Let us turn to an outline of technical manifestations and realisations in health care of integrated care. The integration of healthcare software systems has remained one of the most prominent issues in healthcare software development (Xu et al., 2000:157; Mykkänen, 2003:173; Kuhn and Giuse, 2001). Boochever (2004:16), for instance, underscores that "system integration would provide the platform for improved workflow, patient throughput and patient safety, as well as decreased cost".

An integrated solution is supposed to give the physicians easy access to data from multiple information sources (Winsten, 2000; Friedman, 2001:1529; Tsiknakis et al., 2002:6), thus providing a complete picture of the patient/client's medical history. The multiple information sources are accessed seamlessly from a single point of end-user interaction (Boochever, 2004:16). This avoids that the physician must perform redundant activities, such as specifying the patient identifying information over and over again (Ginneken, 2002:101; Tsiknakis et al., 2002).

The literature on integration of information systems is heavily dominated by presentations of technical mechanisms for integration as solutions for the earlier described problems of fragmentation. There does exist, however, critical voices to the one-sidedness of the above position on integration. Goodhue et al. (1992) have called for a more nuanced approach to integration. Working out a pragmatically based contingency model, they identify conditions under which they argue that the costs (in terms of loss of flexibility, increase in development costs) may outweigh the benefits of integration. Similarly, and more recently, Singletary (2004) surveys practitioners' perceptions of downsides to integration including lock-in with vendors, costs and project risks (see also Markus, 2001). Empirically underpinned case studies (see e.g. Hanseth et al., 2006; Rolland and Monteiro, 2007; Perrow, 1984); demonstrate in more detail the form and implications of the unintended consequences of integration.

# 3. Method

The case study has been conducted at the University Hospital of North Norway (UNN) during January 2004 – May 2005, and has after that been followed up by informal contacts throughout 2006. UNN has approximately 5000 employees, including 450 physicians and 1000 nurses. The hospital has 600 beds of which 450 are somatic and 150 psychiatric. Together with 10 smaller hospitals in North Norway, UNN is administrated by the regional health enterprise Health Region North. Health Region North is responsible for a regional health policy in the northern region, a sound economy and coordination of activities among the 11 hospitals. The health enterprise has also identified information technology as a strategic area, especially related to standardised systems across the hospitals in the region.

In addition to the clinical departments, UNN has 7 laboratories: Clinical Chemistry, Microbiology, Pathology, Clinical pharmacology, Immunology, blood bank and Medical genetics. Together, these laboratories conduct approximately 3 million analyses a year. Clinical Chemistry is the absolutely largest of the laboratories (using number of analyses as a measure), conducting nearly 2 million analyses a year.

The study belongs to an interpretive research approach (Klein and Myers, 1999; Walsham, 1993) and includes participant observations, interviews, document analysis, and informal discussions. The data collection, conducted by the first author, is presented in the table below:

| Participant observations | Participation in 60 project meetings in the HealthSys project from its initial stages (January, 2004) and onwards. This has also resulted in a 70 transcribed documents. |
| --- | --- |
| | The project members participating in these meetings were IT-consultants, physicians, secretaries, bioengineers and nurses. Their number varied from 5 to 16. |
| Informal talks | 50 notes based on informal talks with participants in the HealthSys project and users. |
| Interviews | 12 open-ended interviews with a tape recorder of users of whom eight where physicians. |
| Document archive | 500 emails sent to the HealthSys project and the project document-archive consisting of several 100 documents |

**Table 1 The data collection**

The analysis of the data is based on a hermeneutic approach where a complex whole is understood "from preconceptions about the meanings of its parts and their interrelationships" (Klein and Myers, 1999:71). This implies that the different sources of field data are all taken into consideration in the interpretation process. The method included relatively detailed case write-ups for the sites involved (see for instance Eisenhardt (1989)) followed by an examination of the data for potential analytical themes. Emerging patterns from the data (Schultze, 2000) were attempted categorised to themes in order to make good overviews based on the perspectives chosen. This process was repeated, also involving new theoretical insight.

# 4. Case narrative from the University hospital in Northern Norway

The University Hospital of Northern Norway (UNN) participated for almost eight years in the national and longstanding Medakis project (1996-2003). This project started out as an ambitious, collaborative project between the five Norwegian university hospitals and an international vendor with considerable financial and political backing from the health authorities. The overall goal was to develop a common, all-encompassing EPR for these hospitals covering the needs of all the different health professions in different hospital departments (Ellingsen and Monteiro, 2003). Despite falling significantly short of these expectations, the RecSys EPR has been in operational, increasingly wide-spread, use in the five university hospitals for several years.

The key role of EPRs in Norwegian health care is reiterated regularly in health policy programmes (SHD 1996, 2001). This, however, has not been sufficient to coordinate an integrated and uniform health care. A sweeping health reform in 2002 shifted the ownership of the Norwegian hospitals from the counties over to the Government in an attempt to curb expenditures and poor exploitation of existing resources. The former five health regions were replaced by five regional health enterprises with substantial autonomy, each comprising one of the former university hospitals and several local hospitals.

Increasingly, the users at UNN (especially the physicians), were dissatisfied with the RecSys EPR portfolio. In daily work, they depended on having access to x-ray-descriptions, laboratory results and the EPR. A lack of mutual integration between PAS, EPR and laboratory systems made this situation extremely cumbersome.

The Regional Health Enterprise Northern Norway (here referred to as Health Region North), in a surprising decision in December 2003, exercised its newly gained autonomy to break out of the long-term Medakis collaborative effort. UNN was the only hospital in the northern health region running RecSys EPR, and Health Region North decided to replace this with what the ten other (smaller) hospitals in the region had, namely systems from the vendor HealthSys. To change this was argued to be "obvious". The users perceived RecSys EPR as largely a standalone application. In contrast, HealthSys could presumably offer a complete package including (RIS, PAS, EPR, Laboratory-system and psychiatry). HealthSys promotes their different modules as a complete and integrated solution:

> "The HealthSys solution is based on a common architecture, integrated modules and a common logon-procedure across the different modules."

The HealthSys modules resided in the same database, implying that some registers in the database are shared between the modules. The HealthSys is one out of three vendors on the Norwegian hospital-based EPR market. HealthSys enjoys a 30% market share, mostly at smaller hospitals (Lærum, Ellingsen and Faxvaag, 2001). The HealthSys EPR module comes integrated with a Radiology module, Laboratory module, PAS module and a Psychiatry module.

Health Region North supported the project of replacing RecSys EPR with HealthSys with 10 MNOK (about 1,2 MEUR). Beyond that, UNN was assumed to provide necessary personnel resources. In January 2004, UNN established an implementation group consisting of personnel from UNN and HealthSys.

However, although the HealthSys -system had been successful in the smaller hospitals, the project-group was aware of the risk associated with an IT-project implemented at the region's large university hospital. The project manager underscored that, 'we are about to run a system, which has never been run on this scale before'. A strategy to reduce the risk connected to scale was to tone down the ambition level and just see the implementation of the HealthSys portfolio as just a substitution of the existing hospital portfolio, thus avoiding large organisational changes.

> "No new functionality will be put into use (…) we shall only replace the current PAS and EPR with the HealthSys system (…) there will be no new design of work processes in this phase" (IT-leader).

In spite of this pronounced modest strategy, it was clear to the top management inn UNN that some changes to work- processes were unavoidable, especially in the laboratories. The laboratory module in HealthSys required considerable development in order to provide a satisfactory support for the laboratory work tasks. This was especially a key concern for the Microbiology laboratory. In any case it was seen as crucial to enrol the laboratory in the implementation process both since as many as possible should run a common system an also since this would be the cheapest alternative:

> "A lot of development remains for the Microbiology laboratory (...) but we have to keep on to the vision … and therefore the laboratory must perhaps in a period accept a poorer solution than what they have today" (Hospital manager, SMS)

In addition to the risk of scale, existing integrations were identified as another complication. The old PAS system had over several years been integrated with an increasing amount of smaller systems in the hospital. Completely replacing the old PAS as a big-bang solution was thus considered an extremely risky business as many of the existing systems would stop working as they especially depended on functionality provided by the old PAS-system (searching patients, reimbursement functionality, different codes, reports etc). The HealthSys project experienced that it was difficult to obtain a complete overview over the status of these integrations. The project leader for the technical team put it like this:

> "All those smaller systems imply a chaotic situation. And we don't know who has made the integration, what is integrated and how things are integrated. Moreover we don't know how much [of this functionality] that is actually in use".

As an alternative strategy, it was decided that the old PAS should be kept "alive" side by side with the PAS module in HealthSys. In such a way, all the other systems could be left untouched. It was clearly pointed out that this was just a temporary solution as the IT-leader underscored that 'the old PAS will be shut down at the end of 2005'.

By June 2004, the EPR and the PAS module were implemented. The implementation of laboratory module followed in December 2004.

## 5. Analysis

In the context of health care, the notion of integration has a deeply ambiguous meaning. Moving beyond merely identifying an ambiguity around the notion, the purpose of our analysis is to trace out the career of the vision as played out in the empirical setting under study.

Leaning on the 'career dynamics of organizing visions' (Swanson and Ramiller, 1997), we explore the history of visions on integration in health care, how it transforms in different contexts, but most essentially over time. We focus on how the vision for integrated health information systems – as a result of changes in policy documents, problems and unintended effects surfacing during implementation – go through a career, i.e. it undergoes a sequence of transformations. Our analysis is structured to highlight three sequences or phases of the vision's career. Table 2 summaries this.

| Phase | Notion of integration | Illustration |
|---|---|---|
| Early | One standardised system | "There are 11 hospitals in this region and 10 running the HealthSys portfolio; therefore it is obvious that also UNN should run it" |
| Middle | Single sign-on | "I have three different logon-codes that I have to use on three different systems" |
| Late | Regionalisation of services | "We want to establish shared laboratory services (…) and include similar laboratory functions in common regional units" |

Table 2 The career of the vision of integrated health information systems across three sequences or phases is summarised.

## 5.1 Early phase (2003 – 2004): Before 'integration'

Initially, Health Region North translated the Health authorities' visions of free flow of information, seamlessness and continuity of care into ambitions of a uniform IT portfolio in the northern health region. Basically, this boiled down to insisting that all hospitals in the region should have the *same applications* for each of its functional areas:

> "Within the area of clinical IS support, the strategy for Health Region North is to implement of common, scalable and standard solutions" (Health Region North IT strategy, 2002:29).

> "Clinical results shall be managed in common EPR/PAS systems (…) available for several hospitals and actors in the healthcare sector" (ibid: 2002:28)

Health Region North's ambition was caused by an existing topography in North Norway of 11 hospitals whereof 10 were running the HealthSys system. UNN was the only one not running it. To standardise the IT portfolio into one common system was therefore considered common-sense "obvious", thus black-boxing the issue into an invincible argument (Latour, 1999):

> "There are 11 hospitals in this region and 10 running the HealthSys portfolio; therefore it is obvious that also UNN should run it" (IT manager, Health Region North)

With that, Health Region North evoked the HealthSys system as an appropriate technology (Swanson and Ramiller, 1997:467) for the task.

For Health Region North, a shared regional system meant that upgrades, user requests and general database management could be streamlined and centralised, subsequently followed by significant reduction in expenditures. Health Region North thus grounded the legitimating of a standardised system in broader business concerns of economy and efficiency (Swanson and Ramiller, 1997:461). In line with well-rehearsed arguments for economy of scale, integration qua standardisation of clinical applications was to ensure efficiency gains.

A shared regional system would also legitimize a reorganisation and standardisation of the IT operation procedures. Each of the hospitals had their own associated IT department, which was, according to Health Region North, causing unnecessary expenditures. Implementing one standardised HealthSys system would make it perfectly relevant to also make one regional IT department.

Consequently, a regional IT department (Health Region North ICT), was created in January 2006. The former IT departments in each of the hospitals were decoupled from the hospitals, merged into this common regional IT department consisting of 126 employees. Health Region North ICT is organised directly been under Health Region North control and administration. Key intentions with this establishment were:

> "To contribute to an efficient ICT management through regionalisation (…) it is estimated a minimum profit on 5-10 % in a medium term view, approximately 2-3 years" (Ernst & Young, 2005).

However, the materialisation of the regional IT department really did not fulfil the expectations, at least not in the short term. Bureaucratizing, new expenses and separation from the clinic have been a consequence. An IT-manager at the new IT department expressed it like this:

> "Right now I don't know when the expected efficiency gains might come. In fact, it is more expensive now. Since the IT department is dispersed geographically, we use a lot of resources in reorganisation and coordinating activities across the different locations together with increased expensive flight expenses"

> "Nobody thought about the new expenses related to homogenisation of the infrastructure Right now we have got 100 million NOK to the platform project [common infrastructure]. But we should rather have been funded with 300 mill when the new department was started.

Still, the biggest challenge is definitely the separation from the users, caused by separate organisations:

> "The establishment of an IT department outside the hospitals has caused a big gap between us and them … we don't have the same access to the users in the clinic as before … now there is a border. To some extent … as I see it, the hospitals have forgotten that we are the same people as before when we were part of their organisation … it has been a tough divorce. What's more, the regional IT department represents a new legal organisation, meaning that we must have personnel both among us (key account managers) and in the hospitals that negotiate contracts and service level. This is extremely resource demanding.

## 5.2 Middle phase (2004 – 2005): Integration at the hospital

At UNN, the visions of integration were basically interpreted as "single sign-on". The physicians in the clinic were dissatisfied with their existing IT-portfolio. In daily work, they depended on having access to x-ray-descriptions, laboratory results and the EPR. A lack of mutual integration of their existing PAS, EPR and laboratory systems made this situation difficult as a physician phrases it:

> "I don't have the laboratory results; I don't have the x-ray-description. Instead I have three different logon-codes that I have to use on three different systems [RecSys EPR, Laboratory and RIS] and I have to leave and enter the different systems in turn" (Physician)

HealthSys could presumably offer a complete package including (RIS, PAS, EPR, Laboratory-system and psychiatry). HealthSys promotes their modules as a complete and integrated solution. Implementing the HealthSys portfolio would imply that the users would have to relate to only one system, one interface and one logon procedure:

> "The HealthSys solution is based on a common architecture, integrated modules and a common logon-procedure across the different modules."

Accordingly, when presenting the HealthSys system for the UNN users, the project group traded on the vision's interpretive flexibility (Swanson and Ramiller, 1997:464) and tailored the definition of the HealthSys system exactly to the needs of the users in the clinic – an integrated system with single sign-on. This served to mobilise the physicians at UNN very easily as their most pressing problem would be solved.

The HealthSys portfolio was envisioned as a completely integrated solution for the users. It is timely to inquire *for whom* the resulting work routines became easier – and for whom they implied additional work. The HealthSys system eased the work for the physicians in the clinic, but simultaneously – and unexpectedly – additional work was distributed to the users in the laboratories. The reason for this was that the laboratory users did not use the HealthSys portfolio, and therefore they had to maintain the "integrated solution" partially manually.

The reason for this was that, during and after the implementation, the old PAS was kept "alive" side by side with the PAS module in HealthSys in order to leave all other systems untouched (see figure 2). This strategy required technical integration between the old PAS and HealthSys PAS to maintain the existing information flow and to uniquely identifying patients across these systems. The IT department developed a gateway-program which tracked changes in the HealthSys PAS. Whenever new patients were inserted into the HealthSys system or other changes were made on patient information, this was synchronised with the old PAS.

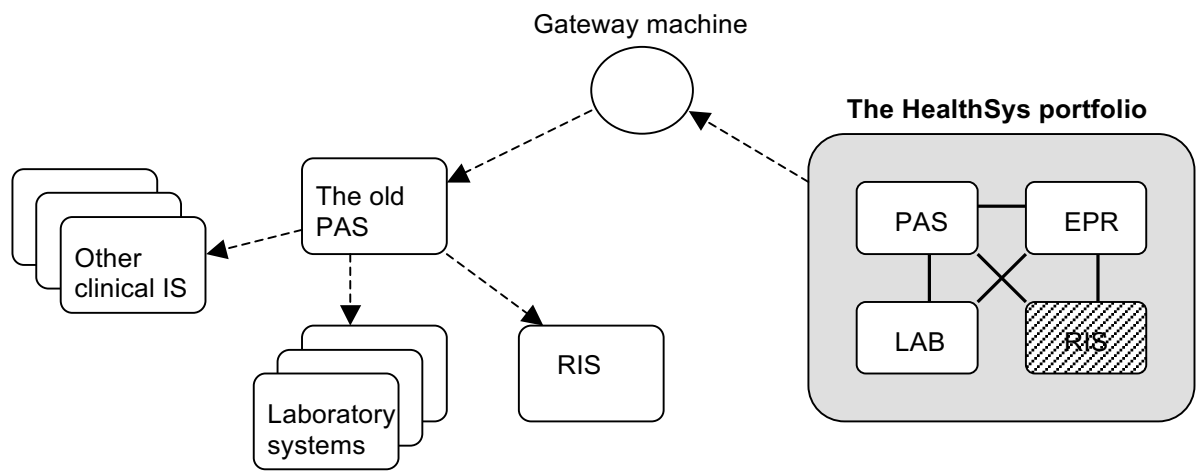

**Figure 2. The Gateway machine connecting the new HealthSys portfolio with the old portfolio, via the old PAS.**

However, this caused additional work for many of the users in the laboratories, as they still used the the old IT portfolio. When a user wanted to create a new sample in one of the old laboratory systems she had first to ensure that the patient existed in the old PAS. First, she had to search for the patient in the old PAS. If this proved futile, which regularly was the case, she would have to log onto the HealthSys system to enter the patient, then wait until the gateway-machine had synchronised patient information with the old PAS. This would easily take 3-4 minutes. Then, finally, she could access the patient in the laboratory system (through its integration with the old PAS).

A smooth information flow between the old IT-portfolio and the HealthSys portfolio was extremely dependent of an up-and-running gateway-machine. However, sometimes the gateway-machine failed to synchronise patients, either because the program had stopped due to system failure or unexpected other reasons, for instance, due to lacking or incomplete data fields. Accordingly, in some situations, when the users had inserted a new patient into the HealthSys system, the patient was never synchronised with the old PAS. At one instance, a laboratory user called the IT-department and complained that she had waited for 2 hours for the synchronisation. In such cases, the users had no choice, but to insert the patients in the old PAS anyway. The synchronisation would eventually become complete when the gateway machine was up and running again and could replace the demographic data written into PAS with data written into the HealthSys system.

In sum, the notion of integration did not imply free flow of information and single sign-on for everybody: predominantly the physicians enjoyed this. In contrast, the laboratory personnel had to do additional work to ensure that the users in the clinic got a complete integrated system. Moreover, the laboratory personnel had to work closely with the IT department, which continuously had to monitor their ad-hoc integration enabled by the improvised gateway machine. Thus, for the laboratories and the IT department, integration became improvisation, monitoring and repair-work (see Ellingsen and Monteiro, 2006 for a more elaborate analysis).

## 5.3 Late phase (2005 – 2007): Regionalised integration

Recently, integration has become embedded in broader political restructuring of Norwegian health care into a stronger regionalisation of services. The notion of integration has thus moved beyond only connecting technical systems to also include coordinated services. As illustrated by Chesbrough and Spohrer (2006) in their research manifesto for services science, integration is now recast into abstract production and consumption of 'services':

> "How can this information of the capabilities of artefacts and organizations be recombined and accelerated in its velocity to create value? How can it be integrated in context to create new and valued services and solutions to customer problems?" (ibid:39)

The vision of coordinated services was picked up by the management of Health Region North to focus on how services should be integrated on a regional level. Practically this has meant trying to implement organisational flexibility and absorbing workload peaks, common bid for tender and standardisation of services. In particular, however, there has been an increased interest in streamlining the laboratory services. Health Region wants a close coordination among the laboratories in the health region, both organisationally and technically. Organisationally, this implies sharing functions, such as a common automated front-end to the laboratories and the inclusion of similar laboratory functions in common organisational units, such as Blood bank North, Clinical Chemistry North and Microbiology North. Technically, this implies to establish common laboratory systems for the hospitals in the region.

However, things did not turn out as expected. When introducing the integrated HealthSys system, the laboratories were expected to replace their systems with the common HealthSys laboratory module in order to comply to ensure tight integration with the rest of HealthSys modules. For the Clinical Chemistry, Immunology and Clinical Pharmacology laboratories, this was an unquestionable argument, thus being in line with Health Region North's ambition on shared services in the health region. The pressure for integration – implying a corresponding pressure for standardisation in the sense of uniformity - was (as we have elaborate on above) embedded in the strategic intentions of Health Region North:

> "For (…) blood bank, pathology and Microbiology we want the same systems and preferably in the same database for each specialised discipline in the whole region" (Director, Health Region North).

The HealthSys laboratory module was up and running for the Clinical Chemistry, Immunology and Clinical Pharmacology laboratories in November 2004. However, the Microbiology laboratory refused to implement the HealthSys laboratory module even if HealthSys promised to put in considerable resources into improving the laboratory module. Despite having a really outdated, existing system, the Microbiology laboratory argued that their complex work routines differed so much from the others. They argued that the HealthSys laboratory module was tailored for simplicity, thus tailored to Clinical Chemistry:

> "The problem is that every vendor making laboratory systems starts out where the production is most intensive and that means Clinical Chemistry, but the problem is that Clinical Chemistry has an incredible simple data structure" (Physician, Microbiology laboratory).

The process inside the Microbiology laboratory is much more complex involving collecting information from various information sources, thus generating highly contextual information. In the HealthSys laboratory module, this information was either missing (for example registers for non-human samples, material, location and antibiotics) or dispersed throughout the system:

> ''In the HealthSys module, generally, the information that we need is very much dispersed in different screens and folders. This requires a lot of clicks orientation and manoeuvres to group things together, such as which bowl, which colony [cultivating] and where the result comes from as traceability is extremely important'' (Physician, Microbiology laboratory)

Accordingly, the ambition of a full-scale regional integrated service for the laboratories has not been possible to obtain.

# 6. Conclusion

Integration in health care is a compelling yet ill-defined notion. This lack of clarity is not accidental, we argue, as 'integration' should be recognised more as a forceful vision (Pfaffenberger, 1998; Swanson and Ramiller, 1997) than a readily definable notion.

A constitutive aspect of visions is their ambiguity. Visions are effective, i.e. able to organise and mobilise political backing, exactly because they allow multiple interpretations. The open-ended nature of integrated system implies that images and metaphors play a particularly productive role as a powerful ally (Boland and Greenberg 1992) for different stakeholders, when implementing instances of the visions.

More importantly than merely noting the ambiguity (or interpretative flexibility), visions qua epistemic objects (Knorr Cetina 2001) are essentially incomplete: they are discovered, changed and extended continuously throughout implementation. This is in part due to the accompanying negotiation among stakeholders, and in part due to the gradual uncovering of new consequences – some intended, others not – over time.

We draw the following implication. The characteristic, ongoing transformations point to the futility of pursuing a completely integrated portfolio as the vision most probably has changed before this stage is ever reached. Instead, stakeholders planning and commissioning integration effort need to be prepared to refocus their efforts in tandem with how the vision changes in order to exercise some influence on how the vision is shaped. We also believe that broad rather than narrow technical evaluations of integration efforts should be conducted. It is not sufficient to evaluate particular instances of integration. Rather, integration needs to be evaluated as a part of a larger career of implementations and transformations in order to get the whole picture of where vision has come from and where it is going.